\RequirePackage{fixltx2e}
\documentclass[english,preprint]{elsarticle}
\usepackage[T1]{fontenc}
\usepackage[latin9]{inputenc}
\usepackage{geometry}
\usepackage{float}
\usepackage{pifont}
\usepackage{natbib}
\usepackage{textcomp}
\usepackage{amsbsy}
\usepackage{amstext}
\usepackage{amssymb}
\usepackage{graphicx}
\usepackage{setspace}


\usepackage{babel}

\begin{document}

\title{Ground state magnetization of conduction electrons in graphene with
	Zeeman effect}

\author[uns,ifisur]{F.~Escudero\corref{cor1}}
\ead{federico.escudero@uns.edu.ar}
\author[uns,ifisur]{J.S.~Ardenghi}
\ead{jsardenhi@gmail.com}
\author[uns,ifisur]{L.~Sourrouille}
\ead{lsourrouille@yahoo.es}
\author[uns,ifisur]{P.~Jasen}
\ead{pvjasen@uns.edu.ar}

\cortext[cor1]{Corresponding author}
\address[uns]{Departamento de Física, Universidad Nacional del Sur, \\
	Av. Alem 1253, B8000CPB, Bahía Blanca, Argentina}

\address[ifisur]{Instituto de Física del Sur (IFISUR, UNS-CONICET), \\
	Av. Alem 1253, B8000CPB, Bahía Blanca, Argentina}

\begin{abstract}
In this work we address the ground state magnetization in graphene,
considering the Zeeman effect and taking into account the conduction
electrons in the long wavelength approximation. We obtain analytical
expressions for the magnetization at $T=0$, where the oscillations
given by the de Haas van Alphen (dHvA) effect are obtained. We find
that the Zeeman effect modifies the magnetization by introducing new
peaks associated with the spin splitting of the Landau levels. These
peaks are very small for typical carrier densities in graphene, but
become prominent for higher densities. The results obtained provide
insight of the way in which the Zeeman effect modifies the magnetization,
which can be useful to control and manipulate the spin degrees of
freedom.
\end{abstract}
\maketitle
\section{Introduction}

Since its experimental isolation in 2004, graphene has become one
of the most studied and promising material in solid state physics
\cite{Geim 1,Geim 2,Zhang,Castro Neto}. Its interesting properties
lie in its 2D hexagonal structure, made of two interpenetrating sublattices
\emph{A} and \emph{B} which behave as a pseudospin degrees of freedom
\cite{Wallace}. Without impurities or defects, the conduction and
valence bands touch at the Fermi energy, with the valence band full
and the conduction band empty in the ground state \cite{Castro Neto}.
Furthermore, in pristine graphene the density of states at the Fermi
energy is zero, and thus the graphene is a semiconductor with zero
band gap, or a semi-metal \cite{Ganhua Lu}. In the long-wave approximation
the dispersion relation is relativistic and the electrons behave as
massless fermions, moving with a Fermi velocity of about c/300 \cite{Goerbig}. 

When a magnetic field is applied to graphene, the discrete Landau
levels are obtained \cite{Kuru}. For a classical electron gas these
levels are equidistant, due to the parabolic dispersion relation.
For a relativistic-like electron gas, as in graphene, the Landau levels
are not equidistant, which is one of the reasons why the Quantum Hall
effect can be observed in graphene at room temperatures \cite{Novoselov,Ardenghi-2,Zhang-1,Gusynin,Zheng,Bolotin}.
Moreover, the Landau levels create an oscillating behavior in the
thermodynamics potentials. It is found that the magnetization oscillates
as a function of the inverse magnetic field, the so called de Haas
van Alphen effect \cite{dAvA,Shoenberg}. The different frequencies
involved in the oscillations are related to the closed orbits that
electrons perform on the Fermi surface \cite{Onsager}. In graphene
it has been predicted that the magnetization oscillates periodically
in a sawtooth pattern, in agreement with the old Peierls prediction
\cite{Sharapov}. In contrast to 2D conventional semiconductors, where
the oscillating center of the magnetization remains exactly at zero,
in graphene the oscillating center has a positive value because the
diamagnetic contribution is half reduced with that in the conventional
semiconductor \cite{Zhen-Guo}.

When we consider the Zeeman effect, the Landau levels for each spin split
introducing a gap. This splitting become relevant when the thermodynamical
properties are considered \cite{Chang,Ardenghi-3}. Indeed, the splitting
affects the filling of the energy states when the internal energy
is calculated, and consequently other related functions such as the
magnetization. In general, the parameters that affect the occupancy of the energy levels
are the electron density $n_{e}$ and the magnetic field. Thus one
can conceive a graphene-like system with its valence band full and
only the conduction band available, in such a way that $n_{e}$ can
be modified. The added electrons could be due to a gate voltage $V_{G}$
applied to the graphene sheet so that $n_{e}$ can be varied as a
function of $V_{G}$. This system may be found useful in the characterization
of spin-filter \cite{Hanson} and spin-polarized currents in two-dimensional
systems \cite{Potok}, which in turn can be used to calculate transport
parameters like charge and spin conductivity \cite{Sinitsyn}. Motivated
by this we have studied the magnetization at $T=0$ in a general graphene-like
system with only the conduction band available and $n_{e}$ variable,
taking into account the Zeeman effect and the way in which the magnetic
oscillations are altered by this effect. Our results include not only
the usual magnetic oscillations but small perturbations given by the
Zeeman effect. We have organized the work as follow: In section 2
we obtain the energy levels of graphene in a magnetic field with Zeeman
effect, and in section 2.1 we study the magnetization and discuss
the results. 

\section{Graphene in magnetic field }

We shall consider the conduction electrons in a graphene system in
the wavelength approximation, which implies low energies such that
$E\ll t\sim3$ eV (where $t$ s the NN hopping amplitude \cite{Castro Neto}).
This gives a relativistic-like dispersion relation $E=\hbar\upsilon_{F}\left|\mathbf{k}\right|$,
where $\upsilon_{F}\sim10^{6}$ m/s is the Fermi velocity. We suppose
that the conduction electrons have an electron density $n_{e}$, which
may be due to an applied gate voltage. The long wavelength approximation
is valid if $n_{e}$ is such that the Fermi energy $E_{F}$ obeys
$E_{F}\ll t\sim3$ eV. With $N$ conduction electrons in an area $A$
we have $n_{e}=N/A$, and the density of states in the long wavelength
approximation is $\rho(E)=g2\pi AE/h^{2}\upsilon_{F}^{2}$,
where $g=4$ takes into account the spin and valley degeneracy. Thus 

\begin{equation}
N=\int_{0}^{E_{F}}\frac{8\pi AE}{h^{2}\upsilon_{F}^{2}} dE.
\end{equation}
Therefore the condition $E_{F}\ll t$ implies

\begin{equation}
n_{e}\ll\frac{4\pi t^{2}}{h^{2}\upsilon_{F}^{2}}.\label{cond ne}
\end{equation}
For $t\sim3$ eV, Eq.(\ref{cond ne}) is satisfied for typical carrier
densities in graphene \cite{Craciun,Nagashio,Hwang,Das Sarma} (about
$n_{e}<10^{12}\mathrm{\;cm^{-2}}$). Then we shall take the regime
$n_{e}\leq0.1\:\mathrm{nm^{-2}}$.

The graphene Hamiltonian in the long wavelength approximation reads\footnote{To find the energies we consider the $K$ valley. }

\begin{equation}
H=\upsilon_{F}(\boldsymbol{\sigma}\cdot\mathbf{p}),\label{Ham-1}
\end{equation}
where $\mathbf{\boldsymbol{\sigma}}=(\sigma_{x},\sigma_{y})$ are
the Pauli matrices, which act in the sublattices $A$ and $B$ of
graphene. Applying a magnetic field, the momentum changes following
the Peierls substitution \cite{Peierls} $\mathbf{p}\rightarrow\mathbf{p}-e\mathbf{A},$
where $\mathbf{A}$ is the vector potential. For a magnetic field
$\mathbf{B}=\left(0,0,B\right)$, in the Landau gauge we have $\mathbf{A}=(-By,0,0)$.
Considering the Zeeman effect \cite{Zeeman}, the term $\mathbf{\boldsymbol{\mu}}\cdot\mathbf{B}=\mu_{B}gBs_{z}/2$
is added to $H$, where $s_{z}=2S_{z}/\hbar$ is the Pauli matrix
acting in the spin state. Therefore Eq.(\ref{Ham-1}) now reads

\begin{equation}
H=\upsilon_{F}\left[\sigma_{x}\left(p_{x}+eBy\right)+\sigma_{y}p_{y}\right]-\mathbf{\boldsymbol{\mu}}\cdot\mathbf{B}.\label{Ham-2}
\end{equation}
Because $H$ only depends on the $y$ coordinate, then we can express
the wave function as $\psi=e^{-ikx}(\begin{array}{cc}
\psi^{A} & \psi^{B}\end{array})$, with $\psi^{A/B}$ depending only on $y$. Replacing $p_{i}=-i\hbar\partial_{i}$
in Eq.(\ref{Ham-2}), the equation $H\psi=E\psi$ becomes

\begin{equation}
\left[v_{F}\left(\sigma_{x}\left(-\hbar k+eBy\right)-i\hbar\sigma_{y}\partial_{y}\right)-\mathbf{\boldsymbol{\mu}}\cdot\mathbf{B}\right]\psi=E\psi.\label{Ham-3}
\end{equation}
Introducing the ladder matrices $\sigma_{\pm}=\sigma_{x}\pm i\sigma_{y}$
and making the change of variable $y'=\left(-\hbar k+eBy\right)/\sqrt{\hbar eB}$
\cite{Ardenghi} we can write Eq.(\ref{Ham-3}) as

\begin{equation}
\left[\upsilon_{F}\sqrt{\hbar eB}\frac{\sigma_{+}}{2}\left(y'-\partial_{y'}\right)+\upsilon_{F}\sqrt{\hbar eB}\frac{\sigma_{-}}{2}\left(y'+\partial_{y'}\right)-\mathbf{\boldsymbol{\mu}}\cdot\mathbf{B}\right]\psi=E\psi.\label{Ham-4}
\end{equation}
This Hamiltonian is identical to the quantum harmonic oscillator Hamiltonian.
Indeed, defining the ladders operators $a^{\dagger}=\left(y'-\partial_{y'}\right)/\sqrt{2}$
and $a=\left(y'+\partial_{y'}\right)/\sqrt{2}$ we have

\begin{equation}
\left[\frac{\hbar\omega_{L}}{2}\left(\sigma_{+}a^{\dagger}+\sigma_{-}a\right)-\hbar\omega_{Z}s_{z}\right]\psi=E\psi,\label{Ham-5}
\end{equation}
where $\omega_{L}=\upsilon_{F}\sqrt{\frac{2eB}{\hbar}}$ and $\omega_{Z}=\mu_{B}gB/(2\hbar)$.
The energies from Eq.(\ref{Ham-5}) can be calculated by writing the
wave function as 

\begin{equation}
\left|\psi\right\rangle =c_{1}\left|n,A,+\right\rangle +c_{2}\left|n-1,B,+\right\rangle +c_{3}\left|n,A,-\right\rangle +c_{4}\left|n-1,B,-\right\rangle ,
\end{equation}
where $\left|+\right\rangle $ and $\left|-\right\rangle $ represent
spin up and down, so that $s_{z}\left|\pm\right\rangle =\pm\left|\pm\right\rangle $.
Then, given that $\sigma_{+}\left|A\right\rangle =0$, $\sigma_{+}\left|B\right\rangle =2\left|A\right\rangle $,
$\sigma_{-}\left|A\right\rangle =2\left|B\right\rangle $, $\sigma_{-}\left|B\right\rangle =0$
and $a^{\dagger}\left|n\right\rangle =\sqrt{n+1}\left|n+1\right\rangle $,
$a\left|n\right\rangle =\sqrt{n}\left|n-1\right\rangle $, solving
Eq.(\ref{Ham-5}) the energies read

\begin{equation}
E_{n,s,l}=l\hbar\omega_{L}\sqrt{n}-s\hbar\omega_{Z},\label{energies}
\end{equation}
where $n=0,1,2\ldots$ is the Landau level index, $s=\pm1$ for spin
up and down and $l=\pm1$ for the valence and conduction band. For
the $K'$ valley the energies are identical to Eq.(\ref{energies}),
so that each state is doubly degenerate. The Zeeman interaction splits
the spin up and down energies, introducing a gap given by $\Delta E=2\hbar\omega_{Z}$.
As in the classical case, the degeneracy of each spin level is given
by $D=2AB/(h/e)=AB/\phi$, where $A$ is the graphene sheet area and
$\phi=h/(2e)$ is half the magnetic unit flux \cite{Goerbig}.\footnote{The factor of 2 in $D$ takes into account the valley degeneracy.} 

To study the ground state magnetization we consider that only the
conduction band is available. The valence band, although full in our
model, would still make a continuous non oscillatory contribution
to the magnetization. Since we are interested only in the magnetic
oscillations and the spin magnetization, we shall omit the valence band and work only with the conduction electrons. Thus
the energies are $\varepsilon_{n}^{s}=\hbar\omega_{L}\sqrt{n}-s\hbar\omega_{Z}$,
where $s=\pm1$ for spin up and down. The internal energy for $N$
electrons can be computed as the sum of the filled Landau levels.
The number of totally filled levels is $q=\left[q_{c}\right],$ where
$q_{c}=N/D$ is the filling factor, and the brackets means the biggest
integer less or equal to $q_{c}$ (the Floor function). We can also
write $N/D=B_{C}/B$ where $B_{C}=n_{e}\phi$ ($n_{e}=N/A)$ is the
critical magnetic field at which the degeneracy $D$ equals the number
of electrons $N.$ 

In order to calculate the internal energy we first have to sort the
energy levels. It may happen that the splitting is such that for a
given Landau level $n$, $\varepsilon_{n+1}^{+}<\varepsilon_{n}^{-}$
, which would mean that the states with energy $\varepsilon_{n+1}^{+}$
are filled before those with energy $\varepsilon_{n}^{-}$. This would
happen if $\hbar\omega_{L}\left(\sqrt{n+1}-\sqrt{n}\right)<2\hbar\omega_{Z}$.
For $q$ levels filled, considering that each state can be occupied
with spin up or down, the condition at which the mixing starts can
be approximated as

\begin{equation}
\hbar\omega_{L}\left(\sqrt{\frac{q}{2}}-\sqrt{\frac{q}{2}-1}\right)<2\hbar\omega_{Z}.\label{B}
\end{equation}
In general this condition depends on the electron density $n_{e}$
because $q=\left[n_{e}\phi/B\right]$. Nevertheless, it can be easily
proved that Eq.(\ref{B}) occur only for electron densities
that do not satisfy Eq.(\ref{cond ne}). Therefore
there is no spin mixing in the long wavelength approximation with
magnetic field. One would have to take in consideration the whole dispersion 
relation of Bloch electrons in graphene \cite{Goerbig} in order to study spin mixing, in which case
the problem becomes increasingly difficult \cite{Plochocka}.  

\subsection{Ground state magnetization }

We call $\xi_{m}$ the decreasing sorted energy levels, $m$ being
the label index. We can write $\xi_{m}=\varepsilon_{m}^{0}-\left(-1\right)^{m}\beta\hbar\omega_{Z}$,
where $\varepsilon_{m}^{0}=\hbar\omega_{L}\left[\frac{m}{2}-\frac{1}{4}\left(1-\left(-1\right)^{m}\right)\right]^{\frac{1}{2}}$
are the Landau levels, written in such a way to ensure that for each
Landau level we take both spins. We introduced a parameter $\beta$
to differentiate the situations without Zeeman effect ($\beta=0$)
and with Zeeman effect ($\beta=1$). 

If we call $\theta=q_{c}-q=N/D-\left[N/D\right]$ the occupancy factor
of the last partially filled Landau level, the ground-state internal
energy reads

\begin{equation}
U=\sum_{m=0}^{q-1}D\xi_{m}+D\theta\xi_{q}.\label{UT}
\end{equation}
Replacing the expression for $\xi_{m}$ we have $U=D\sum_{m=0}^{q-1}\varepsilon_{m}^{0}-\beta D\hbar\omega_{Z}\sum_{m=0}^{q-1}\left(-1\right)^{m}+D\theta\varepsilon_{q}^{0}-\beta D\hbar\omega_{Z}\left(-1\right)^{q}$.
The factor $\sum_{m=0}^{q-1}\left(-1\right)^{m}$ is 0 if $q$ is
even, or 1 if $q$ is odd. Thus we can write $\sum_{m=0}^{q-1}\left(-1\right)^{m}=\left[1-\left(-1\right)^{q}\right]/2$.
Moreover, the term $U_{0}=D\sum_{m=0}^{q-1}\varepsilon_{m}^{0}+D\theta\varepsilon_{q}^{0}$
is the energy without Zeeman effect. Therefore

\begin{equation}
U=U_{0}-\beta\frac{1}{2}D\hbar\omega_{Z}\left[1+\left(-1\right)^{q}\left(2\theta-1\right)\right].\label{Uz}
\end{equation}
The last term in Eq.(\ref{Uz}) is associated with the spin magnetization.
To see this we start from the Pauli magnetization given by 

\begin{equation}
M_{P}=\mu_{B}\left(N_{+}-N_{-}\right),\label{Mp}
\end{equation}
where $N_{+}$ and $N_{-}$ are the total number of spin up and down,
respectively. If $q$ is even, the number of spin up and down states
totally occupied are identical, and the last unfilled state is spin
up. Then $N_{+}-N_{-}=D\theta$ if $q$ is even. On the other hand,
if $q$ is odd there is one unpaired totally filled spin up state
and the last unfilled state is spin down. Then $N_{+}-N_{-}=D-D\theta$
if $q$ is odd. Therefore in general we can write $N_{+}-N_{-}=D\left[1+\left(-1\right)^{q}\left(2\theta-1\right)\right]/2$,
and Eq.(\ref{Mp}) becomes 

\begin{equation}
M_{P}=\mu_{B}\frac{D}{2}\left[1+\left(-1\right)^{q}\left(2\theta-1\right)\right].\label{Mp2}
\end{equation}
Consequently the internal energy of Eq.(\ref{Uz}) becomes

\begin{equation}
U=U_{0}-\beta BM_{P}.\label{Uz2}
\end{equation}
Thus the energy, and related functions such as the magnetization,
are altered by the Zeeman effect through the spin magnetization. The
 magnetization at $T=0$ is $M=-\partial U/\partial B$. From Eq.(\ref{Uz2})
we have

\begin{equation}
M=M_{0}+\beta\left(M_{P}+B\frac{\partial M_{P}}{\partial B}\right),\label{M1}
\end{equation}
where $M_{0}=-\partial U_{0}/\partial B$ is the magnetization without
Zeeman effect. From Eq.(\ref{Mp2}) we get $\partial M_{P}/\partial B=M_{P}/B+\mu_{B}D\left(-1\right)^{q}\partial\theta/\partial B$,
with $\partial\theta/\partial B=-N/(DB)$. Therefore Eq.(\ref{M1})
reads

\begin{equation}
M=M_{0}+\beta\left(2M_{P}-N\mu_{B}\left(-1\right)^{q}\right).\label{MT2}
\end{equation}

In Figure 1 it is shown the magnetization (\ref{MT2}) for an electron
density $n_{e}=0.1\;\mathrm{nm^{-2}}$ and area $A=10\;\mathrm{nm^{2}}$,
for the case without Zeeman effect ($\beta=0$) and the case with
Zeeman effect ($\beta=1$). The magnetization oscillates in agreement
with the de Haas van Alphen effect. Moreover, because $q$ is a periodic
function with $1/B$, therefore $M$ is also a periodic function with
$1/B$, as can be seen in the right of Figure 1. We also notice that
the oscillating center of the magnetization has a positive value,
which means that conduction electrons have a ground state paramagnetism.
This differs substantially with what happens in the conventional semiconductor
2DEG. Thus the results obtained affords an intuitive explanation of
the difference in magnetization between the monolayer graphene and
the conventional semiconductor 2DEG.

\begin{figure}[H]
\includegraphics [scale=0.32] {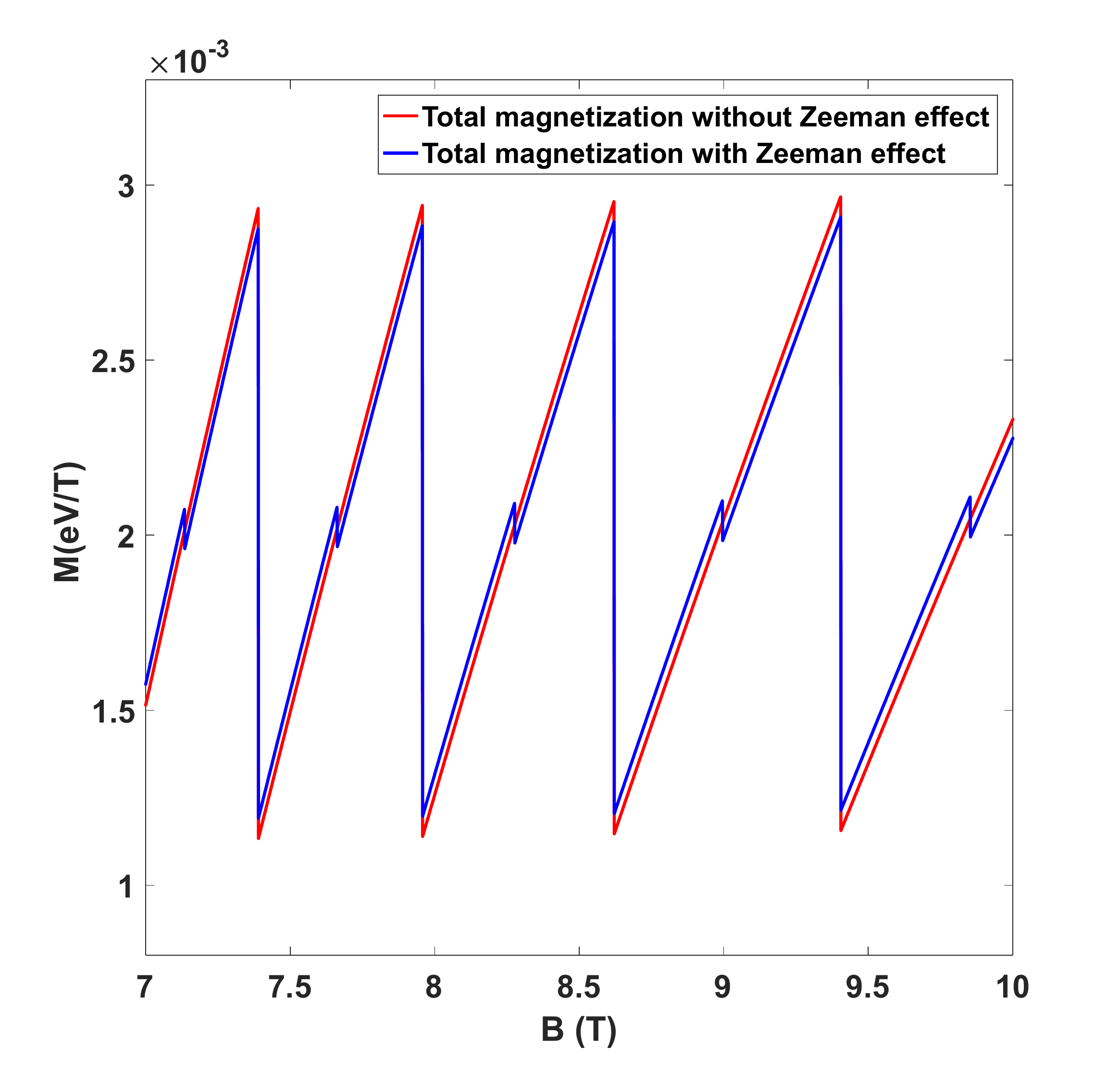}
\includegraphics [scale=0.32] {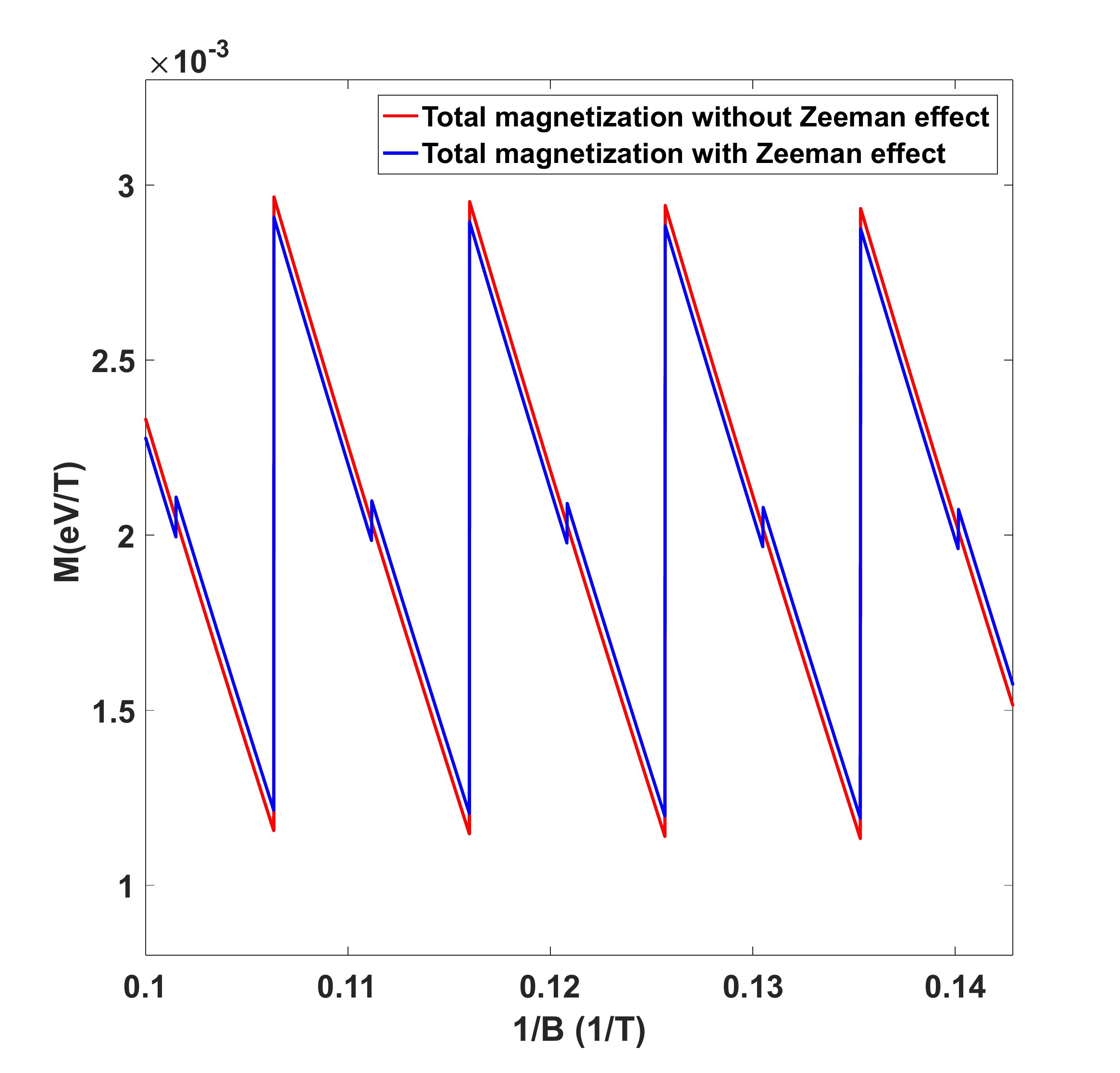}
\caption{Left: Ground state magnetization $M$ as a function of $B$, with
and without Zeeman effect. Right: Ground state magnetization $M$
as a function of $1/B$, with and without Zeeman effect. In both cases
the density of electrons is $n_{e}=0.1\;\mathrm{nm^{-2}}$ and the
area $A=10\;\mathrm{nm^{2}}$.}
\end{figure}

As it can be seen in Figure 1 the Zeeman effect introduces a second
peak in the magnetization. To understand this unusual behavior we
have to analyze in more detail the Eq.(\ref{MT2}). Without Zeeman
effect we have $M_{0}=-\partial U_{0}/\partial B$, with $U_{0}=D\sum_{m=0}^{q-1}\varepsilon_{m}^{0}+D\theta\varepsilon_{q}^{0}$.
Then, given that $\partial D/\partial B=D/B$, $\partial\varepsilon_{m}^{0}/\partial B=(1/2B)(\varepsilon_{m}^{0})$
and $\partial\theta/\partial B=-N/(DB)$, we can write

\begin{equation}
M_{0}=\frac{1}{B}\left(N\varepsilon_{q}^{0}-\frac{3}{2}U_{0}\right),\label{M0}
\end{equation}
and therefore from Eq.(\ref{MT2}) the magnetization with Zeeman effect
becomes

\begin{equation}
M_{Z}=\frac{1}{B}\left(N\varepsilon_{q}^{0}-\frac{3}{2}U_{0}\right)+2M_{P}-N\mu_{B}\left(-1\right)^{q}.\label{Mzeeman}
\end{equation}
From Eq.(\ref{M0}) we see that the peaks appear in $M_{0}$ whenever
$\varepsilon_{q}^{0}$ changes discontinuously, $U_{0}$ being continuous.
This happens only when $q$ changes from odd to even (recall that
$\varepsilon_{m}^{0}=\hbar\omega_{L}\left[\frac{m}{2}-\frac{1}{4}\left(1-\left(-1\right)^{m}\right)\right]^{\frac{1}{2}}$),
which corresponds to a change of Landau level. On the other hand,
Eq.(\ref{Mzeeman}) gives peaks in $M_{z}$ whenever $q$ change because
of the additional factor $N\mu_{B}\left(-1\right)^{q}$; the new peaks
are produced by the change of spin. These results imply that without
Zeeman effect there is a jump in the magnetization only when the last
state changes the Landau level, while with Zeeman effect there is
a jump when the last state changes either its spin or Landau level.
This effect can also be related to fractional filling factors. To
see this consider the energy degeneracy $D_{L}$ of each Landau level
with no Zeeman effect, which can be occupied with spin up or down,
so $D_{L}=2D=2AB/\phi$. Then, $q=\left[N/D\right]=\left[2N/D_{L}\right]=2q_{L}$,
where $q_{L}=\left[N/D_{L}\right]$. For the case with Zeeman effect,
the change of spin is associated with $q$ odd, while the change of
Landau level with $q$ even. In terms of $q_{L}$ this implies that
the peaks in $M$ given by a change of Landau level correspond to
$q_{L}$ integer, whereas the peaks given by change of spin correspond
to $q_{L}$ fractional. In this way we can say that the peaks produced
by the Zeeman effect correspond to fractional filling factors in the
case without Zeeman effect. Such behavior is similar to the Fractional
Quantum Hall effect in graphene \cite{Bolotin,Du,T=000151ke}, where
changes appear in the Hall conductivity for fractional occupancy number 
due to the Coulomb interaction between electrons. 

In the case with Zeeman effect, Figure 1 also shows that the amplitude
of the peaks corresponding to a change of spin is smaller than the
amplitude corresponding to a change of Landau level\footnote{When there is a change of Landau level the spin also changes. But this is different to what we simply call a change of spin, where the spin changes but the Landau level is the same.}. In fact, the
amplitude depends on the density of electrons $n_{e}$, as can be
seen in Eq.(\ref{Mzeeman}). If the peak corresponds to a change of
spin we have $\Delta M^{S}=2N\mu_{B}=2A\mu_{B}n_{e}$, whereas if
it is a change of Landau level, $\Delta M^{L}=\frac{An_{e}}{B}\left[\hbar\omega_{L}\left(\sqrt{\frac{q}{2}}-\sqrt{\frac{q}{2}-1}\right)-2\hbar\omega_{Z}\right]$
with $q$ an even integer. Notice that $\Delta M^{S}$ depends only
on $n_{e}$, while $\Delta M^{L}$ depends on both $n_{e}$ and the
magnetic field $B$. Moreover, the spin splitting appears as a reduction
factor in the amplitude $\Delta M^{L}$, as expected \cite{Shoenberg 2}.
The ratio $\Delta M^{L}/\Delta M^{S}$ is

\begin{equation}
\frac{\Delta M^{L}}{\Delta M^{S}}=\frac{\hbar\omega_{L}}{2\hbar\omega_{Z}}\left(\sqrt{\frac{q}{2}}-\sqrt{\frac{q}{2}-1}\right)-1\qquad\left(q\:\mathrm{even\:integer}\right)\label{deltas}
\end{equation}
where $q=n_{e}\phi/B$ ($\phi=h/2e$). In Figure 2 is plot Eq.(\ref{deltas})
as a function of $n_{e}$, for different values of $q$ even. We see
that for the regime $n_{e}\leq0.1\:\mathrm{nm^{-2}}$ we always have
$\Delta M^{L}>\Delta M^{S}$, but $\Delta M^{L}/\Delta M^{S}$ decreases
as $n_{e}$ increases. For typical carrier densities in graphene,
about $n_{e}<10^{12}\mathrm{\;cm^{-2}}$, we have $\Delta M^{L}\gg\Delta M^{S}$.
This would explain why this phenomenon has not yet been seen in graphene.
Nevertheless, for higher electron densities the effect would be prominent
(see Figure 1). These results are in concordance with \cite{Kishigi},
where the spin splitting appears as a reduction factor in the magnetic
oscillations for 2D normal systems. When the Zeeman splitting becomes
a half of the the Landau level spacing, the amplitude of oscillation
of the fundamental frequency becomes zero. Indeed, Eq.(\ref{deltas})
gives $\Delta M^{L}=0$ for $q=2$ if $\hbar\omega_{Z}=\hbar\omega_{Z}/2$.
Nevertheless, given that $q=n_{e}\phi/B$, this would happen for electron
densities $n_{e}$ that do not satisfy Eq.(\ref{cond ne}).

\begin{figure}[H]
\begin{centering}
\includegraphics [scale=0.32] {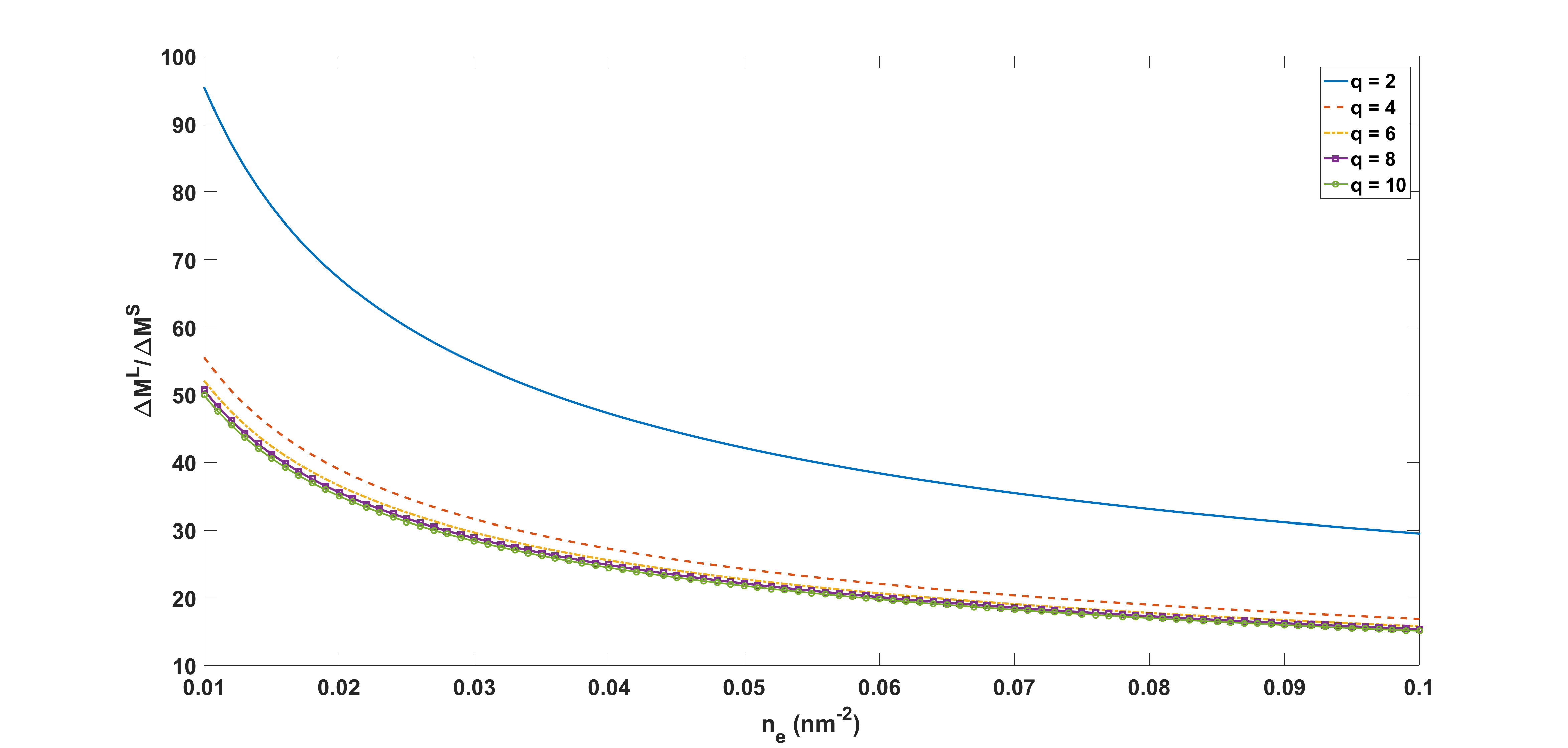}
\end{centering}
\caption{$\Delta M^{L}/\Delta M^{S}$ as a function of $n_{e}$, as given by
Eq.(\ref{deltas}), for different values of $q=n_{e}\phi/B$ ($\phi=h/2e$).}
\end{figure}

From a experimental point of view, the relations of the peaks can
be controlled by the applied electric field that controls the carrier
concentration, whereas the spin polarization lifetime can be controlled
by the applied gate voltage \cite{Xing-Tao An}. This can be useful
to improve the methods for mapping the Fermi surface by taking into
account the Fourier decomposition of the new peaks. 

\section{Conclusions}

We have studied the ground state magnetization of conduction electrons
in graphene with Zeeman effect. We consider only the conduction electrons
in the long wavelength approximation, which was shown to hold for typical 
carrier densities in graphene. We have derived 
analytical expressions for the magnetization at $T=0$, with and without
Zeeman effect. It was shown that the magnetization has peaks whenever
the last energy level changes discontinuously, and its amplitude depends
on the electron density. In the case without Zeeman effect these peaks
appear only when the last Landau level occupied changes. With Zeeman
effect it was shown that new peaks appear in the magnetization, associated
with the spin splitting in the Landau levels. These new peaks occur
whenever the last state changes only its spin, while the Landau level
remains the same. From this we have studied the ratio of amplitudes
between the peaks produced by a change of Landau level ($\Delta M^{L}$)
and the new peaks produced only by a change of spin ($\Delta M^{S}$).
An analytical expression was derived, which shows that $\Delta M^{L}\gg\Delta M^{S}$
for typical carrier densities in graphene, about $n_{e}<10^{12}\mathrm{\;cm^{-2}}$.
Nevertheless, for higher electron densities, about $n_{e}\simeq0.1\;\mathrm{nm^{2}}$,
the effect should become evident. These new findings can be verified
by studying experimentally graphene at very high carrier densities
and perpendicular magnetic field. The predicted effect will hopefully
help the interpretation of magnetization in experiments.

\section{Acknowledgment }

This paper was partially supported by grants of CONICET (Argentina
National Research Council) and Universidad Nacional del Sur (UNS)
and by ANPCyT through PICT 1770, and PIP-CONICET Nos. 114-200901-00272
and 114-200901-00068 research grants, as well as by SGCyT-UNS., J.
S. A. is a member of CONICET., F. E. is a fellow researcher at this
institution.

\end{document}